\documentstyle[pra,aps,twocolumn]{revtex}
\begin{document}
\draft
\title{Optimal compression for mixed signal states}

\author{Micha\l{} Horodecki \cite{poczta1}}

\address{Institute of Theoretical Physics and Astrophysics\\
 University of Gda\'nsk, 80--952 Gda\'nsk, Poland}

\def\dox{\vrule height 4pt width 3pt depth2pt}
\def\hcal{{\cal H}}
\def\ecal{{\cal E}}
\def\ccal{{\cal C}}
\def\qcal{{\cal Q}}
\def\pcal{{\cal P}}
\def\ccal{{\cal C}}

\def\krop{,\ldots,}
\def\trace{{\rm Tr}}
\def\dim{{\rm dim}}
\def\supp{{\rm supp}}
\def\be{\begin{equation}}
\def\ee{\end{equation}}
\def\beq{\begin{eqnarray}}
\def\eeq{\end{eqnarray}}

\def\zrodlo{\hcal^{\otimes N}_Q}
\maketitle
\def\ra{\rangle}
\def\la{\langle}

\begin{abstract}
We consider the problem of  the optimal compression rate in the case
of the source producing mixed signal states within the {\it visible} scheme
(where Alice, who is to compress the signal, can know the identities of
the produced states). We show that a simple strategy based on replacing the
signal states with their {\it extensions} gives {\it optimal} compression.  As a
result we obtain a considerable simplification of the formula for optimal
compression rate within visible scheme.
\end{abstract}



Given a quantum source, what is the minimal number of quantum resources needed
for faithful transmission of the states produced by the source? This
quantum analogue of the problem of data compression \cite{Shannon}
was stated for the first time by Schumacher \cite{Schumacher95}. It has 
been solved for  stationary memoryless sources generating pure states
\cite{Schumacher95,SJ,Barnum}. In general,
the stationary memoryless sources are described by ensemble
$\{p_i,\varrho_i\}$. This means that the source emits system in state
$\varrho_i$ with probability $p_i$ (one can of course generalize it by
considering probability
measure on the set of states). The lack of memory implies that
the state of the sequence of systems emitted by the source is in product state
$\varrho_{i_1}\otimes\ldots\otimes \varrho_{i_N}$ and probability of
emission of such string is product of
suitable probabilities $p_{i_1}\ldots p_{i_N}$. Henceforth we
will deal with  such kind of sources. 

It appears, that for pure signal states $\varrho_i$,  
the minimal number of qubits allowing faithful recovery of the input
states is equal to the von Neumann entropy of the density matrix of ensemble
$\varrho=\sum_ip_i\varrho_i$ \cite{my}. Then the von Neumann entropy has clear
interpretation within purely {\it quantum} communication theory.

In the case of  $\varrho_i$ being impure  the problem is unsolved.
Apart from some
particular cases we do not know much  about optimal compression of
ensembles of mixed states. The scheme that succeeded in the pure states case
can be also applied in the present case, compressing the signal down to the
von Neumann entropy. However, one knows that in many cases it is not optimal
compression \cite{Lo}.
A good candidate for the minimal number of qubits
in this general case could be the Levitin-Holevo function \cite{Holevo} of
ensemble
$I_{LH}=S(\sum_ip_i\varrho_i)-\sum_ip_iS(\varrho_i)$ where $S$ is von Neumann
entropy. Indeed, in this quantity the lost of information
caused by impurity of signal states  is taken into account by
subtracting their mean entropy. As a matter of fact, it has been proven
\cite{hol,oni} that $I_{LH}$ is lower bound for the needed number of qubits.
However a very difficult problem, whether this rate of compression can be
reached remains still unsolved. Additional motivation to consider this problem
is that one would like to know whether $I_{LH}$ has interpretation in terms of
{\it qubits} (so far it has the interpretation in terms of capability of sending
{\it classical} bits via quantum states \cite{Hauslanden,Holevo2,Westmoreland}).

In general, the scheme of compression is as follows. Alice (who is to compress
the signal) waits for a long sequence of the systems generated by the source.
Then she performs some operation on the sequence. Her aim is to decrease the
support of the total density matrix of the ensemble of sequences (the least
Hilbert space the matrix lives on), as the number of needed qubits is equal to
the logarithm of the dimension  of the support. However, she must do it
in a clever way, in order not to disturb the signal too much, so that Bob
would be able to recover it with high fidelity.

There are two basic schemes of  compression. In the first one, called {\it
blind}, Alice does not know the identities of the produces states. Then all
she can do is to apply some quantum operation, independent of
the input states.
If, instead, she knows the identities of the states ({\it visible} scheme)
her operation can be state-dependent, so that she has more possibilities.
Of course, Bob does not know the identities of signal states in either case,
so that his operation is always state-independent.
Thus, in general, the compression could be better within the visible scheme.
The optimal compression rate for pure signal states appears to be
independent of the kind of applied scheme \cite{Barnum}. For mixed state
case, the answer is not known.
In both schemes to obtain the optimal compression rate, one must perform
optimization over Alice and Bob actions satisfying the condition of high
fidelity of transmission. All that must be performed in the limit of 
long sequences. Thus the task is exceedingly difficult.

In one of the attempts to solve the problem, a simple visible protocol of
compression
was proposed \cite{oni2,hol}. Namely, Alice can replace the signal states
\cite{signal}
with their {\it purifications} \cite{purification}, applying then Schumacher
(or Jozsa-Schumacher \cite{SJ})
compression protocol to the resulting ensemble of pure states. More 
generally \cite{hol} Alice can replace signal states with their 
{\it extensions} (by extension of a state $\sigma$ we mean another state
that, partially traced, reproduces
$\sigma$) that are not necessarily pure.
This replacement aims at decreasing the von Neumann entropy $S$ of the
density matrix of the initial ensemble to some lower value $S'$. If it is 
possible, then the subsequent Jozsa-Schumacher (JS) protocol will 
compress the signal at rate $S'$ qubits/message, hence with better performance
than in the case 
of direct application of the protocol, resulting in $S$ qubits/message.

In this paper we consider visible scheme. We prove that the above very simple 
strategy provides {\it optimal} compression rate. More
precisely, to compress the signal optimally, Alice should replace the
sequences of  signal states with their extensions chosen in such a way that 
the von Neumann entropy of the resulting ensemble is minimal.  As a result,
we obtain a considerable simplification
of the  formula for optimal compression rate. The very tedious task
of optimization is now reduced to minimization of the von Neumann entropy
of the ensemble of extensions.

Let us now introduce some notation (the same as in Ref. \cite{hol}).
Suppose that the source generates a system in state $\varrho_i^0$ acting on a
Hilbert space ${\cal H}_{\cal Q}$ with probability $p_i^0$. The produced
ensemble $\ecal_0=\{p_i^0,\varrho_i^0\}$ has the density matrix
$\varrho^0=\sum_ip_i^0 \varrho_i^0$. Denote the product
$\varrho_{i_1}^0\otimes\ldots\otimes \varrho_{i_N}^0$ by $\varrho_i$, where
$i$ stands now for multi-index (to avoid complicated notation we  do not write
the index $N$ explicitly unless necessary).
The corresponding ensemble and state are
denoted by $\ecal$ and $\varrho$ respectively. Now Alice performs a coding
operation over the initial ensemble $\ecal$ ascribing to any input state
$\varrho_i$ a new state $\tilde\varrho_i$. The map $\varrho_i \rightarrow
\tilde\varrho_i=\Lambda_A(\varrho_i)$ is supposed to be a quantum operation
i.e. linear completely positive trace-preserving map
for blind scheme or an arbitrary map - for visible one.
In the latter case
we allow Alice to know which states are generated by the source, so
that she can prepare separately each of the states $\tilde\varrho_i$ for
each $i$.

The new states $\tilde \varrho_i$
represent the compressed signal that is then  flipped onto
the suitable number of qubits determined by the dimension of subspace occupied
by the total state $\tilde\varrho$ of the ensemble and sent through the
noiseless channel to Bob.
Now the states
$\tilde\varrho_i$ are to be decompressed to become close to the initial states
$\varrho_i$. For this purpose Bob performs some established quantum
operation $\Lambda$ which of course does not depend on $i$. Then the resulting
states are $\varrho'_i=\Lambda_B(\tilde\varrho_i)$ and the total
scheme is the following
\begin{eqnarray}
&&\varrho_i \quad
\mathop{\longrightarrow}
\limits_{\Lambda_A}^{{\rm  compression}} \quad
\tilde\varrho_i \quad
\mathop{\longrightarrow}
\limits_{I}^{{\rm noiseless \  channel}} \quad
\tilde\varrho_i  \nonumber\\
&&\mathop{\longrightarrow}
\limits_{\Lambda_B}^{{\rm decompression}} \quad
\varrho'_i,
\end{eqnarray}
where $\varrho_i$ and $\varrho_i'$ act on the Hilbert space
$\zrodlo$ while $\tilde\varrho_i$ on the channel Hilbert space
$H_{\ccal}$. Now we should  determine the measure of quality
of transmission $\varrho_i\rightarrow \varrho_i'$. As one knows, there exist
many different metrics on the set of mixed states. The most common ones
are Hilbert-Schmidt distance $D_{H-S}^2(\varrho,\sigma)=
\trace (\varrho-\sigma)^2$, the one induced by trace norm
$||\varrho-\sigma||=\trace|\varrho-\sigma|$ and the Bures metric
$D_B=2-\sqrt{F(\varrho,\sigma)}$, where the {\it fidelity} $F$
\cite{Jozsa,Uhlmann} is given by
\begin{equation}
F(\varrho,\sigma)=
\left[{\rm Tr}\left(\sqrt{\sqrt{\varrho}\,\sigma\sqrt{\varrho}}\right)\right]^2.
\end{equation}
Instead of Bures metric, one usually uses directly the fidelity. The latter
has an appealing property: if one of the states (say $\varrho$) is pure then
it is of the form
\be
F(\sigma,|\psi\ra\la\psi|)=\la\psi|\sigma|\psi\ra.
\ee
In this case the fidelity has a clear interpretation
as probability that the state $\sigma$
passes the test of being $\psi$. The fidelity was used in the problem of
compression of quantum information  \cite{Schumacher95,SJ,Barnum}, and
it is now an important tool in quantum information theory. However some results
were obtained by use of other measures of quality of transmission. To the
author knowledge, there is no special discrepancy among the results
obtained via different metrics. In fact,  it is yet not clear whether and
to what extent different metrics could lead to non-equivalent conclusions.
In this paper we will
use fidelity, partially due to one of its  properties being especially useful
in the context of the problem of extensions we are dealing  with. Namely
\cite{Jozsa,Uhlmann} the fidelity can be expressed in
the following way
\be
F(\sigma,\varrho)=\max_{\psi} |\langle\psi|\phi\rangle|^2,
\label{overlapy}
\ee
where $\phi$ is arbitrary purification of $\sigma$ and the maximum runs
over all possible purifications of $\varrho$. As we will see further, this
property allows to prove an important lemma.

Consequently, the average fidelity
$\overline{F}(\ecal,\ecal')\equiv\sum_ip_i
F(\varrho_i,\varrho'_i)$ will
indicate us the quality of the process of recovery  of quantum information by
Bob after compression by Alice.
Now, for a fixed source determined by the ensemble $\ecal_0$, one considers
the sequence of compression-decompression pairs $(\Lambda_A,\Lambda_B)$
with the property that
\be
\lim_{N\rightarrow\infty}\overline{F}(\ecal,\ecal')=1
\label{high_fidelity}
\ee
(recall that the pair is implicitly indexed by N). 
Such sequences will be called {\it protocols}.

Define now the quantity $R_P$ characterizing the asymptotic degree
of compression of the initial quantum data at a given protocol $P$ by
\begin{equation}
R_P=\lim_{N\rightarrow\infty}{1\over N}\log \dim\tilde \varrho
\label{compr-p}
\end{equation}
Here $\dim\tilde \varrho$ denotes the dimension of the support of the state
$\tilde\varrho$ given by  the number of its nonzero eigenvalues. The
quantity $\log\dim\tilde\varrho$ has the interpretation of the number of
qubits needed to carry the state $\tilde\varrho$ undisturbed
($\tilde\varrho$ is to be transferred through a noiseless channel).
Actually, only one of the signal sequences $\tilde\varrho_i$ is being 
transmitted at a time. However, it is easy to see, that  
$\dim\tilde\varrho$ is the minimal dimension that guarantees transmission 
of any of the states $\tilde\varrho_i$ without disturbance.

Now, given a class $\pcal$ of protocols, we define the quantity
\begin{equation}
I_{\pcal}=\inf_{P\in\pcal}R_P
\end{equation}
which is equal to the least number of qubits per system needed for
asymptotically faithful transmission of the initial signal states from Alice
to Bob within the considered class of protocols (to be strict one needs
$I_{\pcal}+\delta$ qubits per message, where $\delta$ can be chosen arbitrarily
small). Now, if $\cal P$ is the set of visible protocols, the $I_{\cal P}$ is
called {\it effective information} carried by the ensemble \cite{hol} and is
denoted by $I_{eff}$ (optimal rate within the class of {\it blind} protocols
is called {\it passive} information).
As one can see, the definition of the effective information, even though
physically natural, is very complicated from mathematical point of view. One
must optimize the limit (\ref{compr-p}) over the Alice and Bob actions
keeping satisfied the condition (\ref{high_fidelity}) at the same time.
Moreover, the definition does not give any intuition on how the structure
of ensemble could be related to its effective information content.

Let us now try  to reduce the problem to obtain more transparent form of the effective
information.
 Consider the most general compression-decompression protocol.
Any Bob's operation, as a completely positive trace-preserving map,
amounts to (i)  adding ancilla in some pure
state, (ii) performing unitary transformation over the total system and (iii)
performing partial trace \cite{Kraus}. Now, the two first stages
can be incorporated into
Alice action. Then decompression will amount only to performing partial
trace. Of course, the new protocol will give the same rate of compression as
the previous one, because both stages do not change the dimension of the
support of a state. Thus we can consider  optimal protocol in the following
form
\be
\varrho_i \quad
\mathop{\longrightarrow}
\limits_{\Lambda_A}^{{\rm  compression}} \quad
{\varrho_i'}^{ext} \quad
\mathop{\longrightarrow}
\limits_{\Lambda_B=\trace_{anc}}^{{\rm decompression}} \quad
\varrho'_i
\label{opt-prot}
\ee
where ${\varrho_i'}^{ext}$ are some extensions of the state $\varrho_i'$
and they act on the Hilbert space $\zrodlo\otimes \hcal_{anc}$
(in the following, the  extensions of a state $\sigma$ will be denoted by
$\sigma^{ext}$). Then  the optimal compression rate is given
by  \cite{visible}
\be
I_{eff}= \lim_n {1\over N} \log\dim{\varrho'}^{ext}.
\label{nowy_rate}
\ee

We will need the following lemma.

{\bf Lemma.} Let $\varrho,\varrho'$ act on space $\zrodlo$ and let
${\varrho'}^{ext}$, acting on $\zrodlo\otimes \hcal_{anc}$ be extension of
$\varrho'$.
Then there
exists a state $\varrho^{ext}$
acting on $\hcal^{\otimes n}_Q\otimes \hcal_{anc}$ such that (a)
$\varrho^{ext}$ is
 an extension of $\varrho$ (b)
$F(\varrho^{ext},{\varrho'}^{ext})= F(\varrho,\varrho')$.

{\bf Proof.} Let $\hcal_{ext}=\zrodlo\otimes \hcal_{anc}$ and let $\phi'\in
\hcal_{ext}\otimes \hcal_{pur}$ be purification of ${\varrho'}^{ext}$.
Then it is also purification of the state  $\varrho'$. From the formula
(\ref{overlapy}) we obtain that there exists some purification $\phi$ of
$\varrho$ such that $F(\varrho,\varrho')=|\langle\phi'|\phi\rangle|^2$.
Now we can take $\varrho^{ext}=\trace_{\hcal_{pur}}|\phi\rangle\langle\phi|$.
Using the formula (\ref{overlapy}) once more, we get
$F(\varrho^{ext},{\varrho'}^{ext})\geq
|\langle\phi'|\phi\rangle|^2=F(\varrho,\varrho')$. Since the fidelity does
not decrease under partial trace \cite{wiecej} (this can be easily seen from
 (\ref{overlapy})), we obtain $F(\varrho^{ext},{\varrho'}^{ext})
=F(\varrho,\varrho')$.\dox

Let us now formulate the main result of the paper.

{\bf Theorem.} Let $\ecal^{ext}=\{p_i,\varrho^{ext}_i\}$ with
$\varrho^{ext}_i$
being extensions of the signal states $\varrho_i$; let $\varrho^{ext}$ be the
total density matrix of ensemble $\ecal^{ext}$.
Then the optimal compression rate within the visible scheme is given by
\be
I_{eff}= \lim_{N\rightarrow\infty} {1\over N} \inf S(\varrho^{ext})
\label{optimal}
\ee
where the infimum runs over the set of ensembles $\ecal^{ext}$.

{\bf Remarks.} (1) Since one can choose 
trivial $\hcal_{anc}$ ($\hcal_{anc}=\ccal$),
 $\varrho_i$ itself is the extension of $\varrho_i$, too. (2)~One can show
that the limit on the right hand side of the equality (\ref{optimal}) exists. 
Indeed, it follows from the fact that if a sequence $\{a_n\}$ satisfies 
$a_n\leq kn$ for some $k$ and $a_n+a_m\geq a_{n+m}$ for any $m,n$,
then $a_n/n$ is convergent \cite{BNS}.

{\bf Proof.} To prove that the formula
(\ref{optimal}) is valid, we must first provide the protocol that achieves such
rate, and then show that  the latter is equal to the optimal rate
given by  (\ref{nowy_rate}).

To this end, consider the following concatenated protocol (cf. Ref.\cite{hol})
(call it {\it extension protocol} and denote by $EP$).
Alice replaces the signal state
with the state $\varrho_i^{ext}$, and then applies the JS
protocol \cite{SJ}.
The number of needed qubits per system is now equal to the entropy of the
density matrix of the new ensemble. As JS protocol needs no
decompression \cite{SJ}, Bob has only to perform partial trace to come
back to the original space $\zrodlo$.
Then, the overall scheme is the following
\begin{eqnarray}
&&\varrho_{i_1}\otimes \ldots\otimes \varrho_{i_k}
\mathop{\longrightarrow}\limits^{{\rm Alice's \atop action}}
\varrho_{i_1}^{ext}\otimes \ldots\otimes \varrho_{i_k}^{ext} \quad
\longrightarrow
\nonumber \\
&&\mathop{\longrightarrow}\limits^{{\rm JS\atop compression}}
{\varrho_{i_1 \ldots i_k}'}^{\kern-6mm ext}
\mathop{\longrightarrow}\limits^{{\rm Bob's\atop partial \ trace }}
\varrho_{i_1 \ldots i_k}'
\end{eqnarray}
Here $i_j$'s are multi-indices of length $N$;
$\varrho_{i_1}\otimes \ldots\otimes \varrho_{i_k}$ and
$\varrho_{i_1 \ldots i_k}'$ act on the Hilbert space
$\left(\zrodlo\right)^{\otimes k}$ while
$\tilde\varrho_{i_1}^{ext}\otimes \ldots\otimes \tilde\varrho_{i_k}^{ext}$ and
$\tilde\varrho_{i_1 \ldots i_k}^{ext}$ act on
$\left(\zrodlo\otimes\hcal_{anc}\right)^{\otimes k}$.
The former two states can be obtained from the latter ones by tracing over
the space $ \hcal_{anc}^{\otimes k}$. Now, as it was mentioned,
the fidelity does not decrease under partial trace.
Then (as in Ref. \cite{hol} for different transmission quality  measure)
we obtain that average fidelity  produced by the
composed protocol is greater  than or equal to the one within the
``intermediate'' JS compression protocol.
The latter fidelity tends to one if $N$ is kept fixed and
$k$ tends to infinity  (of course, $N$, although fixed, can be
chosen arbitrarily large). Thus the total protocol satisfies asymptotic
fidelity condition (\ref{high_fidelity}).

Since the JS protocol compresses the signal down to the von Neumann
entropy of the ensemble, the extension protocol has compression rate equal to
\be
R_{EP}= \lim_{N\rightarrow\infty} {1\over N}S(\varrho^{ext})
\ee
Minimizing this expression over all possible extensions  of
signal states we obtain the rate of optimal extension protocol (OEP)
\be
R_{OEP}\equiv \inf _{EP} R_{EP}=\lim_{N\rightarrow \infty} {1\over N}
\inf S(\varrho^{ext}).
\ee
Now we must show that $R_{OEP}\leq I_{eff}$. To this end consider the optimal
protocol of the form (\ref{opt-prot}) so that $I_{eff}$ is given by equation
(\ref{nowy_rate}). Then it suffices to find such extensions
$\varrho_i^{ext}$ that
\be
\lim_{N\rightarrow\infty}{1\over N} S(\varrho^{ext})\leq
\lim_{N\rightarrow\infty}
{1\over N}\log\dim{\varrho'}^{ext}.
\label{nier}
\ee
The suitable extensions are suggested by the lemma. Namely, suppose that
$N$ is large, so that  $F(\ecal,\ecal')>1-\epsilon$ (within the considered
optimal protocol). Then, by the lemma there exist extensions
$\varrho_i^{ext}$ of
$\varrho_i$ such that
$\overline{F}(\ecal^{ext},{\ecal'}^{ext})>1-\epsilon$. Now we can use the
inequality proved in \cite{Barnum2} (which is similar to the Fannes inequality
\cite{Fannes}) saying that for states acting on Hilbert space $\hcal$ we have
\be
|S(\varrho)-S(\varrho')|\leq2\log\dim\hcal \sqrt{1- F(\varrho,\varrho')}+1
\ee
if only $F(\varrho,\varrho')>1-{1\over36}$.
By double concavity of square of $F$ \cite{oni} we obtain in our case
that
\beq
&&{1\over N}|S(\varrho^{ext})-S({\varrho'}^{ext})|\leq\nonumber\\
&&4(\log\dim\hcal_\qcal+{1\over N}\log\dim\hcal_{anc})
 \sqrt{\epsilon}
+{1\over N}
\eeq
One can show \cite{przyg}  that it suffices to consider $\hcal_{anc}$
satisfying $\log\dim\hcal_{anc}\leq 2N\log\dim\hcal_Q$.
Thus the entropy (per system) of the state $\varrho^{ext}$ is
asymptotically
equal to the one of ${\varrho'}^{ext}$. Now,  since $S({\varrho'}^{ext})\leq
\log\dim{\varrho'}^{ext}$ we obtain the inequality (\ref{nier}).
\dox

To summarize, we obtained much simpler formula for optimal compression rate in
visible coding scheme. So far the task was to minimize the support of the
states  under Alice and Bob operations constrained by the asymptotic high
fidelity condition. The latter is very difficult to deal with in the case of
mixed states. The present expression does not involve such constraint, nor it
involves Alice and Bob actions. Now one needs  minimize entropy (which is more
feasible than dealing with dimension of the support) varying over extensions
of the ensemble. Thus the constraints are now much more convenient.
An interesting question arises: Can the optimal compression be achieved by
means of {\it pure} extensions (purifications)? A closely related
question is: Given an ensemble, do there exist such purifications, that
the entropy of the purification ensemble is not greater than the entropy of
the initial one? If it is not the case in general, can it be asymptotically
true for typical sequences of states? Finally, one could ask, whether
the limit in the formula (\ref{optimal}) is really needed. It might be the
case that the minimal entropy could be attained by means of extensions of
single signals $\varrho_i^0$. However, almost everywhere in quantum information
theory the {\it collective} operations are much more powerful than the ones
performed on separate systems. Then it is likely, that collective extensions
of long signal sequences are necessary to obtain optimal compression.

We  believe that the presented result will stimulate to answer these
questions, to find whether the Levitin-Holevo function has 
physical sense in terms of quantum bits, 
and eventually, to resolve the
highly nontrivial problem of compression of quantum information carried
by ensembles of mixed states. 

\begin{acknowledgements}
The author is grateful to Ryszard Horodecki
for stimulating discussion and helpful comments. He also would like to thank 
Chris Fuchs for helpful discussion.
The work is supported by 
Polish Committee for Scientific Research, Contract No. 2P03B 143 17.
\end{acknowledgements}


\begin{references}
\bibitem[*]{poczta1} E-mail address: michalh@iftia.univ.gda.pl
\bibitem{Shannon}
E. Shannon, Bell Syst. Tech. J. {\bf 27}, 379 (1948).
\bibitem{Schumacher95}
B. Schumacher, Phys. Rev. A {\bf 51}, 2738 (1995).
\bibitem{SJ}
R. Jozsa and B. Schumacher, J. Mod. Opt. {\bf 41}, 2343 (1994).
\bibitem{Barnum}
H. Barnum, C. A. Fuchs, R. Jozsa and B. Schumacher, Phys. Rev. A {\bf 54}, 4707
(1996).
\bibitem{my}
This result was generalized to cover the case of unknown parameters of the
source, see M. Horodecki, R. Horodecki and P. Horodecki,
Acta Phys. Slovaca {\bf 48}, 133 (1998);
R. Jozsa, M. Horodecki, P. Horodecki and R. Horodecki, Phys. Rev. Lett.
{\bf 81}, 1714 (1998). See also C. Krattenthaler and P. Slater 
quant-ph/9612043; M. Nielsen PhD thesis, University of New Mexico (1998). 
\bibitem{Lo}
Hoi-Kwong Lo,
Opt. Commun. {\bf 119}, 552 (1995); R. Jozsa, unpublished.
\bibitem{Holevo}
A. S. Holevo, Probl. Peredachi Inform. {\bf 8}, 63 (1973).
\bibitem{hol}
M. Horodecki, Phys. Rev. A {\bf 57}, 3364 (1998).
\bibitem{oni}
H. Barnum, C. Caves, C. A. Fuchs, R. Jozsa and B. Schumacher (unpublished);
C. A. Fuchs and J.~van~de~Graaf, IEEE Trans. IT {\bf 45}, 1216 (1999).
\bibitem{Hauslanden} P. Hauslanden, R. Jozsa, B. Schumacher, M.
Westmoreland and W. K. Wooters, Phys. Rev. A {\bf 54}, 1869 (1996).
\bibitem{Holevo2}
A. S. Holevo, IEEE Trans. IT {\bf 44}, 269 (1998);
also  available as e-print quant-ph/9708046;
\bibitem{Westmoreland}
B. Schumacher and M. Westmoreland Phys. Rev. A {\bf 56}, 131 (1997).
\bibitem{oni2}
H. Barnum, C. Caves, C. Fuchs, R. Jozsa and B. Schumacher (unpublished).
\bibitem{purification}
Recall that the purification
of a state $\varrho$ acting on $\hcal$ is a {\it pure} state $P$ acting on
$\hcal\otimes \hcal'$ such that $\varrho$ is its partial trace over the space
$\hcal'$.
\bibitem{signal}
By signal state we mean the state of sequence of the systems emitted by the
source.
\bibitem{Jozsa}
R. Jozsa, J. Mod. Opt. {\bf 41}, 2315 (1994).
\bibitem{Uhlmann}
A. Uhlmann, Rep. Math. Phys. {\bf 9}, 273 (1976).
\bibitem{Kraus}
K. Kraus, {\it States, Effects and Operations: Fundamental Notions of
Quantum Theory}, Wiley, New York, 1991.
\bibitem{visible} The presented reasoning is also true within the blind scheme.
Then the formula (\ref{nowy_rate}) holds  for passive information, too
(although the involved extensions may be different). 
\bibitem{wiecej} In fact, one has  
$F(\sigma,\sigma')\leq F(\Lambda(\sigma),\Lambda(\sigma'))$ for any
trace-preserving completely positive map $\Lambda$ (see 
H. Barnum, C. M. Caves, C. A. Fuchs, R. Jozsa and B. Schumacher
Phys. Rev. Lett. {\bf 76}, 2818 (1996)).
\bibitem{BNS}
See e.g. H. Barnum, M. Nielsen and 
B. Schumacher Phys. Rev.  A {\bf 57}, 4153 (1998).  
\bibitem{Barnum2}
H. Barnum, J. Smolin, and B. Terhal Phys. Rev. A {\bf 58}, 3496 (1998).
\bibitem{Fannes}
M. Fannes, Commun. Math. Phys. {\bf 31}, 291 (1973); see also M. Ohya and D.
Petz, Quantum Entropy and Its Use, p. 22,  Springer-Verlag 1993.
\bibitem{przyg} M. Horodecki, in preparation.
\end{references}
\end{document}